\documentclass[10pt,conference]{IEEEtran}
\IEEEoverridecommandlockouts

\newcommand{\systemname}{\sc{HFuse}}
\newcommand{\sourcecode}[1]{\texttt{#1}}

\usepackage[figure,vlined,linesnumbered]{algorithm2e}
\usepackage{booktabs}   
\usepackage{subcaption} 
\usepackage{multirow}
\usepackage[frozencache]{minted}
\usepackage{graphicx}
\usepackage{amsmath}
\usepackage[T1]{fontenc}
\usepackage[numbers,comma,sort&compress]{natbib}

\SetKwInOut{Input}{Input}
\SetKwInOut{Output}{Output}

\begin{document}

\title{Automatic Horizontal Fusion for GPU Kernels}         

\author{\IEEEauthorblockN{Ao Li}
\IEEEauthorblockA{University of Toronto\\leo@cs.toronto.edu}
\and
\IEEEauthorblockN{Bojian Zheng}
\IEEEauthorblockA{University of Toronto\\bojian@cs.toronto.edu}
\and 
\IEEEauthorblockN{Gennady Pekhimenko}
\IEEEauthorblockA{University of Toronto\\pekhimenko@cs.toronto.edu}
\and
\IEEEauthorblockN{Fan Long}
\IEEEauthorblockA{University of Toronto\\fanl@cs.toronto.edu}}

\maketitle

\begin{abstract}
We present automatic horizontal fusion, a novel optimization technique that
complements the standard kernel fusion techniques for GPU programs. Unlike the
standard fusion, whose goal is to eliminate intermediate data round trips, our
horizontal fusion technique aims to increase the thread-level parallelism to
hide instruction latencies. We also present {\systemname}, a new source to
source CUDA compiler that implements automatic horizontal fusion. Our
experimental results show that the horizontal fusion can speed up the running
time by 2.5\%-60.8\%. Our results reveal that the horizontal fusion is
especially beneficial for fusing kernels with instructions that require
different kinds of GPU resources (e.g., a memory-intensive kernel and a
compute-intensive kernel).

\end{abstract}




\section{Introduction}

Graphics Processing Units (GPUs) are widely used to speed up deep learning
tasks, scientific computation, and even cryptocurrency mining. Each GPU comes
with dozens to hundreds of processing cores
enabling thousands of threads running in parallel to achieve much higher
computational throughput than a normal CPU~\cite{CUDA}.
Despite of the rapid advancement of the GPU hardware, the applications running
on GPUs are always hunger for more performance. For example, training
state-of-the-art deep learning models like ResNet-50 can take 2 hours on 8
Tesla V100 GPUs~\cite{mlperf}.

To speed up computational tasks running on GPUs, especially for deep learning,
people have developed many optimization techniques at the software
level~\cite{tvm,TASO,glow,nGraph, tensorflow,dlvm}.
Among
these techniques, \emph{Kernel fusion} is a popular and effective
one~\cite{kernel-fusion, persistent-rnns,fusion-2010,automatic-fusion,scalable-kernel-fusion, cublas-fusion}
and it is 
adopted by almost all deep learning frameworks~\cite{tensorflow, mxnet,
PyTorch, dlvm, glow,TASO,nGraph}. In GPU programs,
a large computational task (e.g., training a neural network) is broken down
into multiple \emph{kernels}, each of which corresponds to a small
parallelizable sub-task that will be dispatched to GPUs to execute. The idea of
kernel fusion is to combine two or more kernels into one large but equivalent
kernel to potentially improve the overall performance.

The standard kernel fusion technique in the deep learning frameworks combines
kernels \emph{vertically}. The fused kernel will have the same number of
threads as the two original kernels. Each thread of the fused kernel
sequentially combines the instructions of the corresponding threads of the
original kernels~\cite{tensorflow,TASO, PyTorch,nGraph, glow}. The potential performance advantage is from reducing
expensive data round trips to the GPU device memory --- without the fusion the
first original kernel needs to write its output to the memory for the second
kernel to read.\footnote{For tiny kernels, both vertical and horizontal kernel
fusion also reduces kernel launch overhead.} Therefore the standard kernel
fusion application is typically limited to neighboring kernels in the data
dependency graph, i.e., the output of one kernel is the input of another
kernel.


\subsection{{\systemname}: Horizontal Fusion}

We present a novel optimization technique, \emph {automatic horizontal fusion}.
Unlike the standard fusion that aims to eliminate intermediate data round trip,
our horizontal kernel fusion enables the fused kernel to better utilize GPU
resources and to better hide instruction latencies. The horizontal fusion
complements the standard vertical fusion in its application scenarios ---
horizontally fusing two kernels is beneficial if the two kernels contain
instructions that require different types of GPU resources (e.g., a
memory-intensive kernel and a compute-intensive kernel).

We also present {\systemname}, a source to source CUDA compiler that implements
our automatic horizontal fusion technique. Given the CUDA source code of two
kernels, {\systemname} automatically produces the horizontally fused kernel
that is functionally equivalent to the two but runs potentially faster. In the
horizontally fused kernel, the threads are partitioned into two intervals based
on their thread ids. Each interval corresponds to threads for the computation
of one original kernel. The fused kernel combines the instructions of the
original kernels with branch statements. The branch conditions checks the
current thread id to dispatch the execution to the path of the corresponding
kernel. Because threads of two original kernels coexist in parallel during the
execution, the horizontal fusion exploits the thread-level parallelism. It
enables the thread scheduling hardware (e.g., warp schedulers in NVIDIA GPUs)
to automatically interleave instructions from different kernels to hide
instruction latencies.

One challenge of implementing the automatic horizontal fusion is to handle
synchronization barriers. A typical CUDA barrier stalls the execution of all
threads in a thread block of a kernel until all of the threads reach the
barrier. Because a fused kernel contains threads derived from both of the
original kernels, such barriers from one of the original kernels will impact
the thread execution of another. Another challenge {\systemname} faces is to
identify the best way to partition the thread space of a fused kernel, which is
shared by the instructions of the two original kernels.  Because the partition
scheme determines how the execution of the original kernels co-exists in
GPU, it may significantly impact the performance of the fused
kernel.

To address the first challenge, {\systemname} combines inline PTX assembly
instructions with instrumented branch conditions to implement special barriers
for the thread sets that correspond to original kernels. To
address the second challenge, {\systemname} operates with an automatic
profiling technique. Given the expected input sizes of two original kernels,
{\systemname} will automatically search the best thread space partition.



\subsection{Experimental Results}



We evaluate {\systemname} with 5 deep learning computational kernels extracted
from PyTorch~\cite{PyTorch} and 4 cryptography computational kernels collected
from open source cryptocurrency mining programs~\cite{ethminer, ccminer}. In
total, we apply {\systemname} to fuse 16 pairs of kernels. We compare the
running time of the {\systemname} fused kernel with the native kernels and the
kernels fused in the standard vertical way. Our experimental results show that
the {\systemname} fused kernels run up to 60.8\% faster than the native
kernels; for 7 out of the 16 pairs on 1080Ti GPU and 6 out of the 16 pairs on
V100 GPU, the {\systemname} fused kernels outperform both the vertically fused
kernels and the native kernels. Our results reveal that the speed up of the
fused kernel comes from interleaving different kernel computations to hide
instruction latencies.

Our results also reveal the trade-off between the thread-level and the
block-level parallelism for kernel fusion. One one hand, the horizontal fusion
enables the thread scheduler to interleave instructions with the improved
thread-level parallelism. On the other hand, The fused kernel will require more
register and shared memory resources than individual kernels. If such
additional requirement exceeds a certain breakpoint, it may cause
less thread blocks being scheduled to each core to reduce the block-level
parallelism. One could view {\systemname} as a technique to navigate
this trade-off. See Section~\ref{sec:results:metrics}.

\subsection{Contribution}

This paper makes the following contributions:

\begin{itemize}
\item \textbf{Automatic Horizontal Fusion:} This paper presents automatic
    horizontal fusion, a novel optimization technique that is orthogonal to the standard
        vertical kernel fusion. The horizontal fusion can enable the GPU
        hardware to effectively interleave instructions from two original
        kernels to hide instruction latencies.

    \item \textbf\textbf{{\systemname}:} This paper presents the design and
        implementation of {\systemname}, a novel source to source CUDA compiler
        that implements automatic horizontal fusion.

    \item \textbf{Optimization Scenarios:} This paper identifies the scenarios
        for applying the horizontal fusion technique. Our results show that
        horizontal fusion is mostly beneficial when fusing two kernels with
        instructions that have long latencies and that require different types
        of GPU resources.

\end{itemize}

The remaining of this paper is organized as follows. Section~\ref{sec:overview}
presents an overview of the horizontal kernel fusion technique and a motivating
example of applying {\systemname} to fuse two kernels. Section~\ref{sec:design}
presents the design of {\systemname}.
We then evaluate {\systemname} at
Section~\ref{sec:results}. Section~\ref{sec:related_work} discusses related
work and Section~\ref{sec:conclusion} concludes this paper.

\section{Overview}
\label{sec:overview}

We first introduce background information of GPU architectures that are
important for understanding kernel fusion. We next present an overview of two
kernel fusion techniques, the standard vertical fusion and our horizontal
fusion. We then present an motivating example of applying horizontal fusion
with {\systemname}. In this paper we use the terminology of NVIDIA CUDA
platform~\cite{CUDA} and the architecture parameters of NVIDIA
Pascal~\cite{Pascal} and Volta~\cite{Volta} GPUs. Note that most of the
discussed concepts are generally applicable to other GPU platforms and
architectures.

\subsection{Background}
\label{sec:overview:background}

\noindent \textbf{Kernels, Blocks, and Threads:} Kernels are standalone
computational routines that the CUDA runtime will dispatch to NVIDIA GPUs to
execute in parallel. They are C-like programs that utilize GPU resources
including registers, local shared caches, and the global GPU memory. GPUs are
SIMD processors, so each kernel launch will start multiple \emph {blocks} in
parallel and each block contains multiple threads. The \emph{grid dimension}
(i.e., the number of blocks) and the \emph{block dimension} (i.e., the number
of threads) are typically tunable constants. 
It is a common practice in GPU programming to develop kernels that can work
with different block dimension parameters. This means that changing block
dimensions of the kernels often only influences performance. 
A kernel program can access its
own block id (e.g., \sourcecode{blockIdx.x}) and thread id (e.g.,
\sourcecode{threadIdx.x}) at the runtime to enable its different threads to
potentially process different data.

\begin{figure}[t]
\centering
\includegraphics[width=0.8\columnwidth]{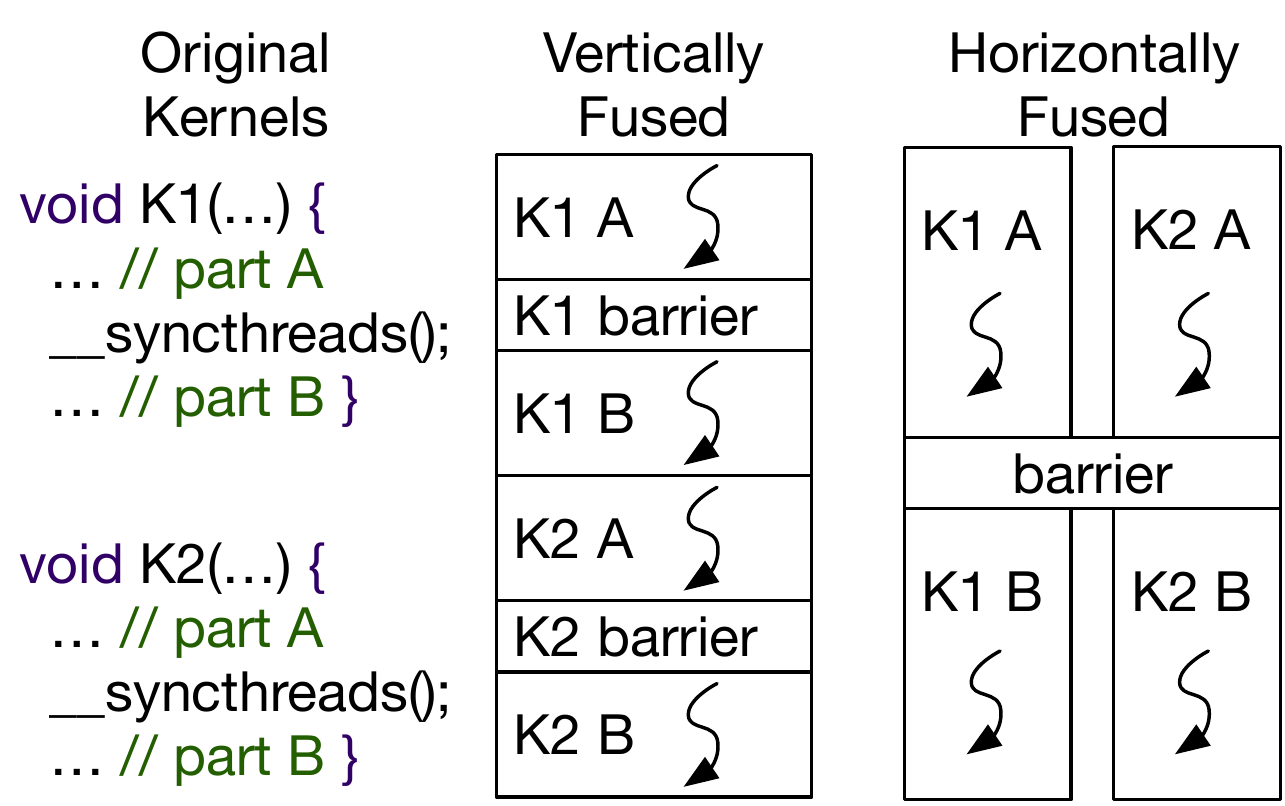}
\caption{Vertical and horizontal kernel fusion.}
\label{fig:fuse}
\end{figure}
\noindent \textbf{Stream Multiprocessor and Occupancy:} When the CUDA runtime
dispatches a kernel to a GPU, the GPU eventually dispatches the blocks of the
kernel to Stream Multiprocessors (SMs) to execute. Each GPU has multiple SMs
depending on its hardware specification. In the Pascal and Volta architectures, each SM
has 64K registers and 96K shared memory cache; each SM can host a maximum of
2048 different threads at the same time; each SM also has multiple CUDA cores
for arithmetic operations and multiple memory controllers for accessing the
global GPU memory.

Because each SM has the fixed amount of resources, it can execute only a
limited amount of blocks in parallel, depending on the kernel resource
requirement. This is called the \emph{occupancy} of a kernel. Generally
speaking, higher occupancy is usually better because it enables the kernel to
exploit the \emph{block-level} parallelism. For example, if a kernel block that
uses 24K shared memory, 512 threads, and 64 registers per thread, a SM can only
execute two blocks in parallel and the registers become the bottleneck. If the
developer optimized the kernel block to use only 32 register per thread, then
the SM can now execute four blocks and the developer doubles the occupancy.

\noindent \textbf{Warps, Warp Scheduler, and Instruction Latency:} In SMs, each
32 consecutive threads form a warp. Threads inside a warp always execute
together in a lock-step fashion and warps are minimum scheduling units in
SMs.\footnote{In the Volta and Turing architectures, warps
do not restrictively execute in the lock-step fashion but warps are still the minimum
scheduling units} The warp scheduler in a SM will select eligible warps to
execute --- an warp is eligible if 1) all data required for its next instruction
is ready, 2) there are idle hardware resources to execute its next instruction
(e.g., idle memory controllers for memory access instructions), and 3) it is
not stalled by barriers.

Because each SM typically has tens of warps executing in parallel, the warp
scheduler may effectively hide \emph{instruction latencies}. If there is a
time-consuming instruction in one warp blocking its execution, warp scheduler
can switch the SM to execute other eligible warps while waiting for the results
of the instruction. Therefore having instructions requesting different hardware
resources in a kernel is beneficial because it tends to increase the number of
eligible warps for the scheduler. It reduces the chance that the SM execution is
completely stalled by instruction latencies.

\begin{figure}[t]
\scriptsize
\input{figs/norm-kernel.tex}
\caption{Normalization kernel.}
\label{fig:norm-kernel}
\end{figure}

\sloppypar{\noindent \textbf{Synchronization Barriers:} The built-in function
\sourcecode{\_\_syncthreads()} in CUDA corresponds to block-wide
synchronization barriers. It is the main way for threads inside a kernel block
to coordinate with each other. An SM will stall the thread execution inside a
block at a block-wide barrier until all threads in the block reaches the
barrier. Note that barriers may significantly limit the capability of warp
schedulers in SMs of hiding instruction latencies, because the schedulers
cannot interleave instructions across the barriers. 
}

\subsection{Kernel Fusion}
\label{sec:overview:fusion}

\noindent \textbf{Vertical Kernel Fusion:} The standard kernel fusion in deep
learning frameworks fuses kernels vertically shown as Figure~\ref{fig:fuse}.
Suppose we have two kernels \sourcecode{K1()} and \sourcecode{K2()} and both of
the kernels have the grid dimension of 512 and the block dimension of 512. The
code of the vertically fused kernel will combine the source code of
\sourcecode{K1()} and \sourcecode{K2()} in order. Therefore the fused kernel
will also has the same gird and block dimensions, but one thread in the fused
kernel will execute the instructions of two original threads, one in
\sourcecode{K1()} and one in \sourcecode{K2()}. The middle part of
Figure~\ref{fig:fuse} shows the execution flow of one thread in the fused
kernel.

The major potential performance advantage of the vertical fusion comes from
eliminating global memory accesses for intermediate results. In this example,
the instructions from \sourcecode{K2()} may directly access the output of
\sourcecode{K1()} without using expensive global memory read instructions. If
some output of \sourcecode{K1()} is only used by \sourcecode{K2()}, the fused
kernel can even eliminate associated global memory write instructions.
Therefore deep learning frameworks typically apply vertical fusion on
neighboring kernels in data dependency graphs.

Note that the vertical fusion may sometime facilitate the instruction
interleaving to hide latency, but such effect is typically minimum due to the
presence of synchronization barriers. The vertically fused kernel will have as
many synchronization barriers as the two original kernels and the warp
scheduler cannot interleave instructions across these barriers.

\noindent \textbf{Horizontal Kernel Fusion:} Unlike the standard kernel fusion,
our horizontal fusion technique creates separate threads for instructions of
different kernels. The right part of Figure~\ref{fig:fuse} presents the
execution flow of the horizontally fused kernel. The fused kernel has the grid
dimension of 512 and the block dimension of 1024. The first 512 threads
correspond to threads for instructions of \sourcecode{K1()} and the remaining
threads correspond to \sourcecode{K2()}. The fused kernel uses branch
statements to check the current thread id to dispatch the thread to execute the
corresponding instructions.

Note that it is possible to partition the thread space of a block unevenly in
the fused kernel, e.g., assigning one kernel 768 threads and another kernel 256
threads. If the block dimensions of the two original kernels are tunable, there
will be multiple ways to fuse the two kernels with different thread space
partition schemes. Which one runs fastest typically depends on the workload of
the original two kernels.

\noindent \textbf{Hypothesis of Horizontal Fusion:}
Our hypothesis of the horizontal fusion is that its thread-level parallelism
will enable the warp scheduler to interleave instructions from different
kernels to hide instruction latencies. It may increase the average eligible warps
on SMs to improve the overall performance. If our hypothesis is true, then the
horizontal fusion will be mostly beneficial for fusing kernels that use
different kinds of instructions and kernels that are memory intensive (because
memory instructions have long latencies). Our experimental results in
Section~\ref{sec:results} validate our hypothesis.

\subsection{Motivating Example}
\label{sec:overview:example}

\begin{figure}[t]
\footnotesize
\input{figs/hist-kernel.tex}
\caption{Histogram kernel.}
\label{fig:histogram-kernel}
\end{figure}

\begin{figure}[t]
\footnotesize
\input{figs/fused-kernel.tex}
\caption{{\systemname} fused kernel.}
\label{fig:fused-kernel}
\end{figure}

We next present an example of using {\systemname} to horizontally fuse two deep
learning kernels. Figure~\ref{fig:norm-kernel} shows the simplified code
snippet of \sourcecode{batch\_norm\_collect\_statistics()}, a CUDA kernel that
computes the mean and variance of an input tensor for normalization. We
extracted this kernel source code from the PyTorch framework~\cite{PyTorch} and
this kernel is used by ResNet~\cite{resnet}. The kernel in
Figure~\ref{fig:norm-kernel} uses intra-warp shuffles~\cite{shuffles} to speed
up its computation. It can operate with a tunable block dimension size as long
as the size is a multiple of 32. Each thread first computes the partial results
of the mean and the variance from the corresponding entries of the tensor with
the loop at lines 10-15. The kernel then uses intra-warp operations to
aggregate the partial results of each warp (consecutive 32 threads) at lines
16-22. It then writes the partially aggregated results of the 16 warps to the
shared memory at lines 26-29 and further aggregates these partial results to
produce the output at lines 33-46.

Figure~\ref{fig:histogram-kernel} shows the simplified code snippet of
\sourcecode{kernelHistogram1D()}, a tensor analysis kernel in PyTorch to
generate histograms over values in an input tensor. Because investigating
tensor value distributions at hidden layers is a common practice for developers
to tune model parameters, this kernel could be invoked during the training of
the ResNet model together with the kernel in Figure~\ref{fig:norm-kernel}.
\sourcecode{kernelHistogram1D()} uses the shared memory array
\sourcecode{my\_smem} at lines 2-3 to count the appearances of tensor values in
different ranges. This kernel also operates a tunable block dimension size. It
initializes the shared counters at lines 6-9. It then iterates the tensor
values to atomically increment the shared counters at lines 12-17 and finally
merges the shared counter results with the global counter output at lines
21-25.

Given the two kernels in Figures \ref{fig:norm-kernel} and
\ref{fig:histogram-kernel} as the input, {\systemname} horizontally combines
them to generate a faster fused kernel shown as Figure~\ref{fig:fused-kernel}
with the following steps.


\noindent \textbf{Generate Prologue:} {\systemname} first generates the
prologue for the fused kernel shown as lines 2-23 in
Figure~\ref{fig:fused-kernel}. The fused kernel has 1024 threads per block. The
first 896 threads correspond to the first input kernel (e.g.,
\sourcecode{batch\_norm\_collect\_statistics()}), while the remaining 128
threads correspond to the second input kernel (e.g.,
\sourcecode{kernelHistogram1D()}). The prologue checks the current thread id
and maps it back to the thread ids of the original kernels, storing them into
the variables \sourcecode{threadIdx\_x}, \sourcecode{threadIdx\_y}, and
\sourcecode{threadIdx\_z}. It also sets variables like \sourcecode{blockDim\_x}
to the original input kernel dimensions. The prologue finally will include all
variable declarations from the two input kernels. {\systemname} properly
renames these local variables to make sure each of them has a fresh name.

\noindent \textbf{Transform Original Kernels:} {\systemname} then transforms
the original two kernels. Lines 37-40 in Figure~\ref{fig:fused-kernel} present
the translated code of the first part of \sourcecode{kernelHistogram1D()}.
{\systemname} replaces the built-in special values with the corresponding
defined variables in the prologue (e.g., replaces \sourcecode{threadIdx.x} with
\sourcecode{threadIdx\_x} and \sourcecode{blockDim.x} with
\sourcecode{blockDim\_x}). {\systemname} then add additional branch statements
to check the current thread id at lines 25 and 36. The branches will skip the
execution of the statements of one kernel if the current thread is in the
thread range of the other kernel.

\noindent \textbf{Replace Synchronization Barriers:} 
\sourcecode{\_\_syncthreads()} will break the original kernel semantics in the
fused kernel, because it will attempt to synchronization all threads in the
fused kernel, which include threads for both of the original kernels. To
preserve the original kernel semantics, {\systemname} replaces synchronization
barriers in the original kernels with inlined PTX assembly
\sourcecode{bar.sync} instructions at lines 29, 32, 42, and 45 in
Figure~\ref{fig:fused-kernel}. Note that the second parameter of
\sourcecode{bar.sync} denotes the number of threads participating the
barrier~\cite{ptx}. {\systemname} passes 896 for this parameter at lines 29 and
32 and passes 128 at lines 42 and 45. Combining with the inserted branch
statements at lines 25 and 36, these \sourcecode{bar.sync} instructions will
create the desired partial barriers that only synchronize threads wihtin the
corresponding thread ranges of each original kernel.

\noindent \textbf{Profile Different Configurations:} Because both of the two
original kernels support tunable block dimensions, there are multiple ways to
partition the thread space of the fused kernel. For example, it is possible to
partition the thread space evenly among two kernels. Additionally, the fused
kernel will use more registers than any of the two input kernels and high
register usage may lower the occupancy. Enforcing a register bound in CUDA may
improve the performance of the fused kernel. 

{\systemname} automatically profiles possible configuration combinations. For
Pascal 1080Ti GPU and the default workload of these two original kernels in our
experiments, {\systemname} outputs the kernel in Figure~\ref{fig:fused-kernel}
as the fastest fused kernel and restricts the register usage to 32 per thread.
The kernel in Figure~\ref{fig:fused-kernel} runs 53.4\% faster than
individually executing two kernels in Figures \ref{fig:norm-kernel} and
\ref{fig:histogram-kernel} on 1080Ti. For Volta V100 GPU, the fastest fused
kernel partitions the thread space differently. It assigns 768 threads instead
of 896 threads for the first kernel and the remaining 256 threads to the second
kernel. It runs 15.8\% faster than individually executing two kernels on V100.

\section{Design}
\label{sec:design}

We next present the design of {\systemname}. 
In this section, we
represent a kernel as a list of CUDA statements. Macros are preprocessed,
function calls are all inlined, and local variable declarations are lifted to
the top of the function. We will discuss these preprocessing steps in
Section~\ref{sec:impl}. In pseudo-codes, we use double quotations to denote
CUDA statements. For simplicity, in this section we assume that the CUDA
kernels have only one block sub-dimension, i.e., \sourcecode{blockDim.y} and
\sourcecode{blockDim.z} are one. It is straightforward to extend our algorithm
to cover kernels with more than one block sub-dimensions. 

\subsection{Generate Fused Kernel}

\begin{algorithm}[t]
  \SetEndCharOfAlgoLine{}

\Input{$K_1$ and $K_2$ are two input kernels. $d_1$ and $d_2$ are the block dimensions of $K_1$ and $K_2$.}
\Output{A fused kernel $F$}

\underline{function} $Generate(K_1, K_2, d_1, d_2):$\\

Initialize $F$ with local variable declarations from $K_1$ and $K_2$ and
extract non declaration statements as $S_1$ and $S_2$.

Append $``\text{tid=threadIdx.x;}$
$\text{tid\_1=threadIdx.x;}$
$\text{tid\_2=threadIdx.x-$d_1$;}$
$\text{size\_1=$d_1$;}$
$\text{size\_2=$d_2$;}"$ to $F$

Replace ``threadIdx.x'' and ``blockDim.x" in $S_1$ and $S_2$ with ``tid\_1'' and ``size\_1" or ``tid\_2" and ``size\_2" accordingly.

Replace ``\_\_syncthreads()'' in $S_1$ with the inlined PTX ``bar.sync 1, $d_1$;"

Replace ``\_\_syncthreads()'' in $S_2$ with the inlined PTX ``bar.sync 2, $d_2$;"

Append ``\text{if (threadIdx.x >= $d_1$) goto $l_1$;}'' to $F$

Mark the end of $S_1$ with the label $l_1$

Append $S_1$ to $F$

Append ``\text{if (threadIdx.x < $d_1$) goto $l_2$;}'' to $F$

Mark the end of $S_2$ with the label $l_2$

Append $S_2$ to $F$

\Return $F$

  \caption{An algorithm generates the fused kernel.}
\label{fig:generate}
\end{algorithm}

Figure~\ref{fig:generate} presents the pseudo-code of $\mathrm{Generate()}$.
Given two input kernels $K_1$ and $K_2$ together with their block dimensions
$d_1$ and $d_2$, $\mathrm{Generate()}$ returns the horizontally fused kernel
$F$. 

\noindent \textbf{Generate Prologue:} The pseudo-code in
Figure~\ref{fig:generate} first copies the local variable declarations from the
two input kernels to the fused kernels at line 2. It properly renames them to
make sure that the local variables do not have conflict names.  At line 3 the
pseudo-code defines and initializes a set of special variables,
\sourcecode{tid\_1} and \sourcecode{tid\_2} for storing the original thread id
of the two input kernels as well as \sourcecode{size\_1} and
\sourcecode{size\_2} for storing the original block dimension of the two
kernels. 

\noindent \textbf{Replace Built-in Variables:} The pseudo-code at lines 4 then
replaces ``\sourcecode{threadIdx.x}'' and ``\sourcecode{blockDim.x}" with the
corresponding defined variables in the prologue. This is because in the fused
kernel, these built-in values will refer to the fused kernel not the original
kernel. This replacement preserves the semantics of the statements in the
original kernels. 

\noindent \textbf{Replace Synchronization Barriers:} \sourcecode{\_\_syncthreads()}
in CUDA implements a barrier for all threads in a block. In the fused kernel,
the instructions from the two input kernels are running concurrently in
different threads of a block, so {\systemname} needs to replace
\sourcecode{\_\_syncthreads()} with partial barriers only for the
threads of the corresponding input kernel. 

Fortunately, the inlined PTX instruction \sourcecode{bar.sync} can support
partial barrier~\cite{ptx}. The first parameter of \sourcecode{bar.sync} is a
constant from 0 to 15 denoting the barrier id. The second parameter of
\sourcecode{bar.sync} is a constant denoting the number of threads
participating the barrier. Internally, the GPU hardware maintains a counter to
track how many threads have reached the barrier. When sufficient threads have
reached the barrier, they are allowed to progress. The pseudo-code at lines
5-6 replaces \sourcecode{\_\_syncthreads()} with \sourcecode{bar.sync} PTX
instructions. These instructions pass the barrier id one for barriers in the
first original kernel and two for barriers in the second kernel. They also pass
the original block dimension as the second parameter to implement the desired
partial barriers. When combined with the branch guards inserted at lines 7 and
10, these \sourcecode{bar.sync} instructions will only wait for threads from
their own original kernels instead of for all threads. The fused kernel
therefore has synchronization barriers at equivalent places for the equivalent
sets of threads as the original two kernels.

\noindent \textbf{Append Guarded Statements:} The pseudo-code finally appends
the translated statements of two input kernels into the fused kernel at lines
7-12. Before appending the statements of each kernel, {\systemname} will insert
an if statement to check the current thread index at lines 7 and 10. In the
fused kernel, the threads in the index range of $[0, d_1)$ correspond to the
first input kernel, while the threads in the index range of $[d_1, d_1 + d_2)$
correspond to the second input kernel. If the index is outside the range of the
corresponding input kernel, it will skip the statements from the kernel. 

\subsection{Search Fusion Configuration}

\begin{algorithm}[t]
  \SetEndCharOfAlgoLine{}
\Input{$K_1$ and $K_2$ are two different kernels. $d_0$ is the desired block dimension of the fused kernel.} 
\Output{A fused kernel $F*$ and the register bound $r*$ for launching the kernel.}

\underline{function} $Main(K_1, K_2, d_0):$\\

$t^* \leftarrow \infty$

$F^* \leftarrow \emptyset$

$r^* \leftarrow \bot$

$d_1 \leftarrow 128$

\While {$d_1 < d_0$} {
    
    $F \leftarrow \mathrm{Generate}(K_1, K_2, d_1, d_0 - d_1)$

    $t \leftarrow \text{Profile the running time of $F$}$

    \If {$t < t^*$} {
        $t^*\leftarrow t$

        $F^*\leftarrow F$

        $r^* \leftarrow \bot$
    }

    $b_1 \leftarrow \frac{\mathrm{SMNRegs}}{d_1 * \mathrm{NRegs}(S_1)}$

    $b_2 \leftarrow \frac{\mathrm{SMNRegs}}{d_2 * \mathrm{NRegs}(S_2)}$

    $b_0 \leftarrow \mathrm{min}(\mathrm{min}(b_1, b_2), \frac{\mathrm{SMShMem}}{\mathrm{ShMem}(F)}, \frac{\mathrm{SMNThreads}}{d_0})$

    $r_0 \leftarrow \frac{\mathrm{SMNRegs}}{b_0 * d_0}$

    $t \leftarrow \text{Profile $F$ with the register bound $r_0$}$

    \If {$t < t^*$} {
        $t^*\leftarrow t$

        $F^*\leftarrow F$

        $r^* \leftarrow r_0$
    }

    $d_1 \leftarrow d_1 + 128$
}

\Return $F^*, r^*$

  \caption{A search algorithm finds the best fusion configuration.}
\label{fig:search}
\end{algorithm}

Figure~\ref{fig:search} presents the pseudo-code of our main algorithm to
search for the best fusion configuration. Given statements from two kernels
$S_1$ and $S_2$ and the desired block dimension of the fused kernel $d_0$, the
algorithm produces a horizontally fused kernel kernel $F^*$ as its output. 

\noindent \textbf{Thread Space Partition:} The pseudo-code uses a loop at lines
6-22 to search for the best thread space partition. At each iteration, it tries
a different block dimension for the first kernel (i.e., $d_1$), generates the
fused kernel at line 7, and profiles the running time of the fused kernel
twice, once without any register bound at line 8 and once with a calculated
register bound at line 17. At lines 10-12 and 19-21, the pseudo-code records
the fastest fused kernel together with its configuration. Note that
{\systemname} searches the block dimension of the first kernel at a granularity
of 128, because using an irregular block dimension often breaks memory access
patterns and causes CUDA kernels to run slower. 

\noindent \textbf{Limit Register Usage for Occupancy: }
The fused kernel may require more registers than each of the original two
kernels. This additional register requirement may lower the occupancy, each SM
will be able to execute less blocks concurrently due to the available total
registers per SM. In practice, the CUDA compiler can enforce a bound to limit
the number of registers used in a compiled kernel. Excessive registers will be
spilled into the global GPU memory. It is therefore possible to recover the
occupancy loss at the cost of introducing expensive memory instructions.

The pseudo-code in Figure~\ref{fig:search} automatically explores this
trade-off with profiling. For each different thread space partition,
{\systemname} will attempt to compile the fused kernel twice with different
configurations, one without the register bound and one with it. When the
algorithm sets the bound, it computes the bound $r_0$ at lines 13-16. Note
that $\mathrm{SMNRegs}$ is the number of registers per SM (64K for Pascal and
Volta GPUs), $\mathrm{SMShMem}$ is the shared memory size per SM (96K for
Pascal and Volta GPUs), $\mathrm{SMNThreads}$ is the maximum number of
concurrent threads per SM (2048 for Pascal and Volta GPUs), $b_1$ and $b_2$ are
the numbers of concurrent active block while launching the original two
kernels, $\mathrm{ShMem}()$ denotes the used shared memory size of a kernel,
and $\mathrm{NRegs}()$ denotes the number of used registers of a kernel. The
intuition of this bound is to make the fused kernel to run as many blocks per
SM as the two input kernels, unless the occupancy is otherwise bounded by the
number of threads or the shared memory usage.

\subsection{Implementation}
\label{sec:impl}

We implemented {\systemname} based on the front-end CUDA parser of the LLVM
Clang framework~\cite{Clang-cuda}. The front-end parser converts the CUDA source
code into abstract syntax trees (ASTs) and {\systemname} traverses ASTs to
implement the algorithm we described in Section~\ref{sec:design}.
For each input kernel file, our implementation uses Clang to pre-process all
macros and included headers. We also use the built-in functionalities from the
Clang front-end to inline all function calls in the input kernels.
{\systemname} does not support recursive function calls. Note that recursion is
extremely rare in GPU kernels, because GPUs are very inefficient at handling it
as well. In our experiments, we do not encounter any kernel that contains
recursive calls.

Additionally, {\systemname} traverses the AST of the input kernel to locate all
local variable declarations. It renames each local variable to make sure that
they will not cause name conflicts in the fused kernel. It also lifts their
declarations to the start of the kernel. If the declaration of a local variable
is associated with initialization assignments, it will still lift the
declaration but create corresponding new assignment statements at the original
location of the declaration. {\systemname} lifts local variable declarations
because it instruments goto statements into the fused kernel and CUDA may not
allow goto statements to jump over local variable declarations.


\section{Experimental Results}
\label{sec:results}

\begin{figure*}[t!]
  \begin{center}
  \includegraphics[width=2\columnwidth]{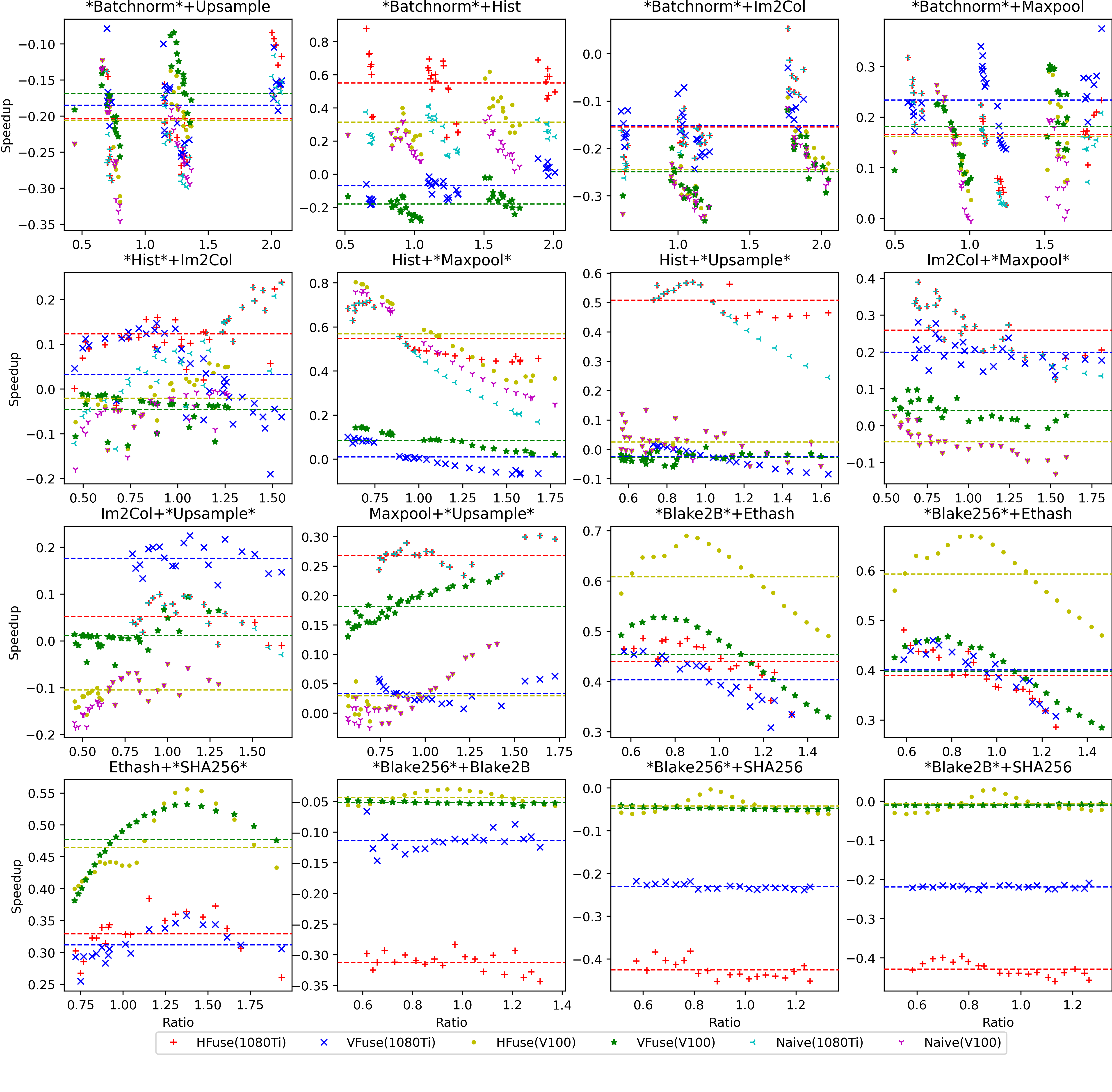}
  \end{center}
  \vspace{-7mm}
  \caption{Kernel execution time speedup.}
  \label{fig:kernel-speedup}
\end{figure*}
We next evaluate {\systemname} with 5 deep learning computational kernels and 4
cryptography computational kernels.
The goal of this evaluation is to answer
the following questions:

\begin{itemize}
  \item How effective is {\systemname}? How do the horizontally fused kernels
      compare with native kernels and the vertically fused kernels in
        performance?
  \item Why do the horizontal fused kernels run faster?
  \item What is the right scenario to apply the horizontal fusion?
  \item How much improvement does the automatic profiling technique have on
      fusing kernels with barriers?
\end{itemize}

\subsection{Methodology}\label{sec:results:methodology}

\noindent\textbf{Benchmark Kernels:} We collect 9 GPU kernels including
5 deep learning computational kernels and 4 cryptography computational kernels:
Maxpool applies a 2D maxpooling over an input matrix;
Batchnorm collects the batch mean and variance a 2D input matrix, which
will then be used for normalization;
Upsample applies a 2D bilinear upsampling over a input matrix;
Im2Col rearranges the input image blocks into columns;
Hist computes the histogram of an input matrix.
These kernels have been widely used in AI models such as ResNet~\cite{resnet}, BigGAN~\cite{biggan},
and UVC~\cite{uvc}.
Ethash is a memory intensive hash function
used by Ethereum~\cite{ethereum} for its proof of work mining. SHA256, Blake256, and Blake2B are three
computational intensive hash functions used for
the proof-of-work of several cryptocurrencies.

All deep learning computational kernels are extracted from
PyTorch~\cite{PyTorch}. Ethash is extracted from ethminer~\cite{ethminer}, and
the rest three cryptography computational kernels are collected from
ccminer~\cite{ccminer}. The 5 deep learning kernels form 10 possible benchmark
pairs, while the 4 crypto kernels form 6 possible benchmark pairs. Note that
all deep learning kernels support tunable block dimensions while crypto kernels
do not.

\noindent\textbf{Apply {\systemname}:}
For each benchmark pair, we apply {\systemname} to horizontally fuse the two
kernels.
We run the fused kernel and measure its running time. For comparison, we
measure the running time of launching the original kernels individually via
parallel CUDA streams. We implement the standard vertical fusion and compares
its running time with {\systemname} as well. To evaluate the effect of our
profiling techniques, we also run a version of {\systemname} that evenly
partition the thread space for two kernels without profiling. Note that because
crypto kernels do not support tunable block dimensions, {\systemname} always
evenly partition the space instead.
We use nvprof, a profiling tool provided by CUDA toolkit, to collect the
performance data of each kernel. In all experiments we take into account two
generations of NVIDIA GPU cards: GeForce GTX 1080 Ti graphic card based on
Pascal, and Tesla V100 graphic card based on Volta.

\noindent \textbf{Run Different Workload:} All benchmark kernels can operate
with variable workload. Deep learning kernels can process inputs with different
sizes, while cryptography kernels can run iterations to compute multiple
hashes. The speed up of any fusion technique depends on the execution time ratio
of two input kernels, i.e., fusing two kernels with similar execution time will
be typically more beneficial. To understand how horizontal fusion works in
different workload ratio, we run our experiments with different input sizes for
each kernel. For each benchmark pair, we will report the speed up under
different execution time ratios of the two original kernels.

\noindent\textbf{Execution Time Measurement:} Since it may take a while for the
GPU performance to stabilize, we launch a dummy kernel on the GPU for about 500
millisecond before launching any experimental kernels. For each pair of
kernels, we record elapsed time after the first kernel launches and before the
second kernel finishes with nvprof as the native execution time. 

\begin{figure*}[t]
\small
  \begin{center}
  \begin{tabular}{ |c|c|c|c|c| }
    \hline
    {\textbf{Kernel}} & \textbf{Execution Time (ms)} & \textbf{Issue Slot Utilization (\%)} &
    \textbf{MemInst Stall (\%)} & \textbf{Occupancy (\%)} \\
\hline
Im2Col &1.92 / 1.69 &87.18 / 63.81 &27.5 / 38.2 & 48.0 / 48.1 \\
\hline
Maxpool &1.93 / 1.97 &7.99 / 8.55 &95.2 / 97.2 & 89.5 / 92.2 \\
\hline
Upsample &1.72 / 2.41 &34.32 / 23.29 &77.8 / 81.3 & 48.3 / 49.7 \\
\hline
Hist &1.70 / 1.90 &14.46 / 50.70 &1.4 / 7.3 & 99.0 / 74.3 \\
\hline
Batchnorm &2.15 / 1.90 &61.83 / 63.27 &52.2 / 60.3 & 96.2 / 98.1 \\
\hline
\hline
Blake256 &38.43 / 37.33 &91.01 / 53.22 &1.3 / 0.0 & 48.9 / 48.9 \\
\hline
Blake2B &39.20 / 39.40 &90.15 / 52.44 &1.7 / 0.0 & 49.1 / 48.9 \\
\hline
SHA256 &42.88 / 39.69 &65.62 / 49.07 &0.0 / 0.0 & 30.6 / 24.6 \\
\hline
Ethash &46.01 / 37.23 &10.88 / 4.17 &96.1 / 96.6 & 36.7 / 18.2 \\
\hline
  \end{tabular}
  \end{center}
  \caption{Metrics of individual kernels.}
  \label{fig:kernel-metrics-individual}
\end{figure*}

\noindent\textbf{Performance Analysis:
}For each benchmark kernel, we select a representative input size so that the
execution time ratios of the ten benchmark pairs are close to one. F To
understand the effect of register bounds, we make {\systemname} to generate two
fused kernels, with and without register bounds. For each variant of the fused
kernels under the representative input size, we use nvprof to collect four
metrics besides its execution time:
\begin{itemize}
  \item \textbf{Issue Slot Utilization:} Percentage of issue slots that issued
  at least one instruction. The streaming multiprocessor is stalled because of
  instruction latencies.
  \item \textbf{MemInst Stall:} Percentage of stalls caused by waiting for
  memory instructions.
  \item \textbf{Occupancy:} Ratio of the average active wraps per active cycle to
  the theoretically number of warps supported on a multiprocessor.
  \item \textbf{Elapsed Cycles:} Elapsed clocks of the launched kernel.
\end{itemize}

We also computes the average issue slot utilization while executing two native kernels
using the following formula:
$$
I_{k1+k2}=\frac{I_{k1} * C_{k1} + I_{k2} * C_{k2}}{C_{k1} + C_{k2}}
$$

where $I_k$ and $C_k$ are the issue slot utilization and the elapsed cycles
of kernel $k$.

\subsection{Performance Results}


Figure~\ref{fig:kernel-speedup} shows the kernel execution time speedup with
respect to the native execution of 16 pairs of kenels. In each subplot,  the
x-axis represents the ratios of execution time of two kernels; the y-axis
represents the speedup of the fused kernel with resepct to the native
execution. Each subplot has four kinds of markers which represent standard
fusion (VFuse) and horizontal fusion (HFuse) on two different GPU generations
(1080Ti and V100). For deep learning kernels, we also include two additional
kinds of markers to represent horizontal fusion without thread space profiling
(Naive) on the two GPU generations.

For each benchmark pair, we change the input size of one benchmark kernel
(marked with ``*'' in the pair name in Figure~\ref{fig:kernel-speedup}) to
obtain results on different execution time ratios of the two kernels. Different
marks of the same kind correspond to experimental results of different input
sizes. Each subplot in Figure~\ref{fig:kernel-speedup} also draws four
horizontal lines in the corresponding color to represent the average speedup
of the fused kernel across different execution ratio data points. Note that the
execution time of Batchnorm changes non-continuously as its input size changes,
so the marks in the four pairs involving Batchnorm appear in clusters.


Our results highlight the effectiveness of our automatic horizontal fusion
technique across two different domains. For 5 out of the 10 deep learning cases
(*Batchnorm*+Hist, *Batchnorm*+Maxpool, Hist+*Maxpool*, Hist+*Upsample*, and
Maxpool+*Upsample*) and for 3 out of the 6 crypto cases (pairs with Ethash),
the {\systemname} fused kernel outperforms the native execution across
different execution time ratios on both 1080Ti and V100. For these cases, the
{\systemname} marks are almost always on the positive side of the y-axis. The
average speedup of {\systemname} over different execution time ratios on these
9 cases are 12.4\%-55.1\% on 1080Ti and 2.5\%-60.8\% on V100.
For 8 cases on 1080Ti and 5 cases on V100 (out of the total 16 cases), the
{\systemname} fused kernel on average over different execution time ratios
outperforms both the standard fusion and the native execution.

We observe that the {\systemname} fused kernels have more significant speedup
on those cases where one of the original kernels is memory intensive. This is
because the horizontal fusion interleaves memory instructions of such a kernel
with other instructions to hide latencies of these expensive memory
instructions. Note that both kernel fusion techniques perform badly on
*Blake256*+Blake2B, *Blake256*+SHA256, and *Blake2B*+SHA256, because those
cryptography computational kernels require similar computational resources and
fusing such kernels together will bring little benefits but harm the
occupancies.



Our results also show the importance of the thread space partition. The
automatic profiling technique enables {\systemname} to better fuse kernels that
have different execution times. For all deep learning cases except
*Batchnorm*+Im2Col, the thread space profiling technique is able to find a
thread space partition scheme that performs better than the naive approach for
some execution time ratio. Better partitioning the thread space will make
threads for the two original kernels to co-exist longer so that the warp
scheduler can better interleave their instructions.

\subsection{Kernel Metrics Results}\label{sec:results:metrics}

\begin{figure*}[t!]
\small
  \begin{center}
  \begin{tabular}{ |c|c|c|c|c|c|c| }
    \hline
    \multirow{2}{*}{\textbf{Pairs}} & \multirow{2}{*}{\textbf{Type}} & \multirow{2}{*}{\textbf{Speedup (\%)}} & \multicolumn{2}{c|}{\textbf{Issue Slot Utilization (\%)}}  & \textbf{MemInst} & \multirow{2}{*}{ \textbf{Occupancy (\%)}}  \\
    \cline{4-5}
    & &  & \textbf{{\systemname}} & \textbf{Native} &  \textbf{Stall (\%)} &   \\
    \hline
\multirow{2}{*}{Batchnorm+Upsample} &N-RegCap &\textcolor{red}{-35.0} / \textcolor{red}{-37.4}  & 32.05 / 25.93 & \multirow{2}{*}{52.29 / 41.69} &67.2 / 76.1 & 42.9 / 42.9 \\
\cline{2-4} \cline{6-7}
 &RegCap &\textcolor{red}{-23.4} / \textcolor{red}{-28.9}  & 38.35 / 29.73 &  &75.0 / 80.7 & 87.1 / 83.2 \\
\hline
\multirow{2}{*}{Batchnorm+Hist} &N-RegCap &\textcolor{green}{51.2} / \textcolor{red}{-28.5}  & 55.89 / 33.43 & \multirow{2}{*}{40.55 / 57.22} &45.6 / 49.7 & 90.7 / 43.0 \\
\cline{2-4} \cline{6-7}
 &RegCap &\textcolor{green}{53.4} / \textcolor{green}{15.8}  & 56.74 / 53.33 &  &46.1 / 56.3 & 91.1 / 96.0 \\
\hline
\multirow{2}{*}{Batchnorm+Im2Col} &N-RegCap &\textcolor{red}{-31.1} / \textcolor{red}{-42.8}  & 44.43 / 31.55 & \multirow{2}{*}{73.65 / 64.81} &50.8 / 67.6 & 37.3 / 42.0 \\
\cline{2-4} \cline{6-7}
 &RegCap &\textcolor{red}{-12.7} / \textcolor{red}{-31.1}  & 58.13 / 41.55 &  &63.4 / 74.5 & 92.0 / 85.3 \\
\hline
\multirow{2}{*}{Batchnorm+Maxpool} &N-RegCap &\textcolor{green}{7.8} / \textcolor{green}{3.4}  & 32.58 / 28.41 & \multirow{2}{*}{35.73 / 35.28} &67.3 / 78.2 & 64.1 / 72.7 \\
\cline{2-4} \cline{6-7}
 &RegCap &\textcolor{green}{7.8} / \textcolor{green}{3.4}  & 32.58 / 32.20 &  &67.5 / 78.2 & 64.0 / 72.7 \\
\hline
\multirow{2}{*}{Hist+Im2Col} &N-RegCap &\textcolor{green}{12.2} / \textcolor{red}{-17.3}  & 60.34 / 40.69 & \multirow{2}{*}{51.90 / 58.03} &20.0 / 31.6 & 38.1 / 37.4 \\
\cline{2-4} \cline{6-7}
 &RegCap &\textcolor{green}{11.3} / \textcolor{red}{-0.1}  & 60.09 / 51.32 &  &19.8 / 46.1 & 38.3 / 78.6 \\
\hline
\multirow{2}{*}{Hist+Maxpool} &N-RegCap &\textcolor{green}{52.5} / \textcolor{green}{56.6}  & 19.07 / 36.25 & \multirow{2}{*}{11.05 / 28.58} &26.7 / 43.5 & 67.5 / 59.0 \\
\cline{2-4} \cline{6-7}
 &RegCap &\textcolor{green}{53.4} / \textcolor{green}{57.1}  & 19.10 / 39.07 &  &25.0 / 43.0 & 67.7 / 57.5 \\
\hline
\multirow{2}{*}{Hist+Upsample} &N-RegCap &\textcolor{green}{4.4} / \textcolor{red}{-5.7}  & 30.17 / 35.01 & \multirow{2}{*}{26.87 / 35.70} &40.8 / 41.2 & 38.5 / 43.5 \\
\cline{2-4} \cline{6-7}
 &RegCap &\textcolor{green}{51.4} / \textcolor{green}{5.7}  & 41.20 / 36.37 &  &48.0 / 57.0 & 77.6 / 82.6 \\
\hline
\multirow{2}{*}{Im2Col+Maxpool} &N-RegCap &\textcolor{green}{7.0} / \textcolor{red}{-12.0}  & 51.52 / 30.47 & \multirow{2}{*}{45.95 / 34.05} &54.7 / 69.3 & 32.6 / 32.5 \\
\cline{2-4} \cline{6-7}
 &RegCap &\textcolor{green}{25.3} / \textcolor{red}{-7.5}  & 57.87 / 33.87 &  &62.5 / 74.4 & 63.5 / 58.4 \\
\hline
\multirow{2}{*}{Im2Col+Upsample} &N-RegCap &\textcolor{green}{5.4} / \textcolor{red}{-10.8}  & 71.92 / 36.72 & \multirow{2}{*}{64.76 / 41.11} &43.0 / 72.2 & 42.7 / 44.9 \\
\cline{2-4} \cline{6-7}
 &RegCap &\textcolor{red}{-24.1} / \textcolor{red}{-45.5}  & 49.50 / 24.24 &  &73.6 / 78.9 & 73.7 / 74.0 \\
\hline
\multirow{2}{*}{Maxpool+Upsample} &N-RegCap &\textcolor{red}{-1.6} / \textcolor{red}{-3.4}  & 23.39 / 16.18 & \multirow{2}{*}{22.47 / 17.00} &79.3 / 86.4 & 30.0 / 33.1 \\
\cline{2-4} \cline{6-7}
 &RegCap &\textcolor{green}{29.4} / \textcolor{green}{1.1}  & 30.32 / 17.97 &  &81.0 / 88.3 & 60.9 / 62.3 \\
\hline
\multirow{2}{*}{Blake2B+Ethash} &N-RegCap &\textcolor{green}{15.9} / \textcolor{green}{30.1}  & 58.93 / 36.73 & \multirow{2}{*}{47.39 / 28.29} &22.8 / 25.5 & 15.8 / 8.6 \\
\cline{2-4} \cline{6-7}
 &RegCap &\textcolor{green}{42.9} / \textcolor{green}{65.8}  & 70.08 / 46.85 &  &19.8 / 23.9 & 29.0 / 29.2 \\
\hline
\multirow{2}{*}{Blake256+Ethash} &N-RegCap &\textcolor{green}{17.0} / \textcolor{green}{30.3}  & 57.89 / 37.05 & \multirow{2}{*}{47.49 / 28.46} &19.5 / 26.5 & 16.1 / 8.6 \\
\cline{2-4} \cline{6-7}
 &RegCap &\textcolor{green}{47.4} / \textcolor{green}{64.7}  & 71.41 / 46.79 &  &17.6 / 24.7 & 29.3 / 29.3 \\
\hline
\multirow{2}{*}{Ethash+SHA256} &N-RegCap &\textcolor{green}{8.8} / \textcolor{green}{37.0}  & 39.25 / 36.62 & \multirow{2}{*}{36.97 / 26.81} &10.3 / 26.8 & 15.7 / 8.8 \\
\cline{2-4} \cline{6-7}
 &RegCap &\textcolor{green}{35.1} / \textcolor{green}{44.1}  & 50.51 / 39.37 &  &18.4 / 16.5 & 28.8 / 28.8 \\
\hline
\multirow{2}{*}{Blake256+Blake2B} &N-RegCap &\textcolor{red}{-26.5} / \textcolor{red}{-2.7}  & 66.08 / 51.34 & \multirow{2}{*}{90.58 / 52.82} &2.3 / 0.0 & 37.5 / 36.8 \\
\cline{2-4} \cline{6-7}
 &RegCap &\textcolor{red}{-96.5} / \textcolor{red}{-96.1}  & 3.60 / 3.31 &  &72.0 / 62.4 & 98.4 / 96.0 \\
\hline
\multirow{2}{*}{Blake256+SHA256} &N-RegCap &\textcolor{red}{-44.3} / \textcolor{red}{-1.0}  & 41.13 / 50.22 & \multirow{2}{*}{77.81 / 51.11} &0.9 / 0.0 & 22.7 / 24.4 \\
\cline{2-4} \cline{6-7}
 &RegCap &\textcolor{red}{-51.2} / \textcolor{red}{-37.4}  & 42.57 / 34.32 &  &43.0 / 7.7 & 56.2 / 51.6 \\
\hline
\multirow{2}{*}{Blake2B+SHA256} &N-RegCap &\textcolor{red}{-42.9} / \textcolor{green}{2.8}  & 41.40 / 50.26 & \multirow{2}{*}{77.49 / 50.74} &0.8 / 0.0 & 22.7 / 24.5 \\
\cline{2-4} \cline{6-7}
 &RegCap &\textcolor{red}{-50.9} / \textcolor{red}{-31.7}  & 38.27 / 35.30 &  &48.0 / 7.6 & 54.9 / 50.6 \\
 \hline
  \end{tabular}
  \end{center}
  \caption{Metrics of {\systemname} fused kernels.}
  \label{fig:kernel-metrics}
\end{figure*}

To understand in what scenario {\systemname} performs best, we collect the
performance metrics of the original kernels and the {\systemname} fused kernel
variants under a representative workload in which the execution time of
benchmark kernels is close to each other.
Figure~\ref{fig:kernel-metrics-individual} shows the results of individual
kernels. Each row corresponds to the metrics of one kernel.
Figure~\ref{fig:kernel-metrics} shows the results of the {\systemname} fused
kernels. Each entry in Figure~\ref{fig:kernel-metrics-individual} and
Figure~\ref{fig:kernel-metrics} is of the form ``X / Y", where X is the result
for 1080Ti GPU and Y is the result for V100 GPU.
In order to understand some key factors that influence the performance
of the fused kernels. We collect metrics for kernels both with register bound
(RegCap) and without register bound (N-RegCap).

The third column in Figure~\ref{fig:kernel-metrics} shows the speedup of the
fused kernel against the native execution. The fourth column presents the
instruction issue slot utilization for the fused kernels and the fifth column
presents the average instruction issue slot utilization computed from the
metrics of two individual kernels (from
Figure~\ref{fig:kernel-metrics-individual}). The issue slot utilization denotes
the percentage of GPU cycles that at least one warp is active for a kernel
(that SMs are not stalled due to instruction latencies). The last two columns
present the percentages of the stalls caused by memory instructions and the
achieved occupancies of the kernels.

\noindent \textbf{Issue Slot Utilization:} Our results indicate that
{\systemname} is effective because horizontal fusion interleaves instructions
to hide the instruction latencies. For all cases, a fused kernel runs faster
than the native execution if the fused kernel has a higher issue slot
utilization. On one hand, the fused kernel will have a much better performance
if two kernels uses two different computational resources. For example, Ethash
is a memory intensive kernel and Blake256 is a compute intensive kernel. As
shown in Figure~\ref{fig:kernel-metrics-individual}, the percentage of stalls
caused by waiting for memory instructions of Ethash is 96.1\% on 1080Ti GPU.
The percentage of Blake256 is only 1.3\%. Therefore, the issue slot utilization
of the fused kernel of Ethash and Blake256 is 23.9\% higher than the native
execution. The fused kernel hides high latency of the memory instructions in
Ethash by interleaving computation instructions from Blake256. On the other
hand, fusing two compute-intensive kernels is not very beneficial, as shown by
the Blake256+Blake2B, Blake256+SHA256, and Blake2B+SHA256 cases.

\noindent \textbf{Thread-level v.s. Block-level Parallelism:} The horizontal
fusion may lower the occupancy, which is another key factor that influence the
performance. Occupancy indicates the ratio of the average active wraps per
active cycle to the theoretically number of wraps supported on a Stream
Multiprocessor (SM). If the number of blocks which can execute concurrently on
a streaming processor is low, the occupancy of the kernel will also be low
because there are not enough eligible warps to be launched. The horizontal
fusion may increase the number of registers per thread of the fused kernel,
which may limit the maximum number of active blocks on an SM.

Therefore one could view the horizontal fusion as a technique to navigate the
inherent trade-off between the thread-level parallelism of interleaving
instructions from more threads and the block-level parallelism of running more
blocks per SM. Our results show that it is often beneficial to apply horizontal
fusion to gain thread-level parallelism even at the cost of block-level
parallelism. For cases including Batchnorm+Maxpool and Hist+Maxpool, the fused
kernels have lower occupancies on 1080Ti and V100 GPUs than both of the
corresponding original kernels but they run faster.

\noindent \textbf{Register Bound:} The register bound may recover the occupancy
loss at the cost of additional memory instructions for spilled registers. Our
results show that the fused kernel with the register bound may perform better
than the kernel without it. For example, Hist+Upsample only achieves 38.5\%
occupancy without a bound, but it achieves 77.6\% occupancy with the bound on
1080Ti GPU. Because of the large improvement of the occupancy, for this case
the version with the register bound runs significantly faster. We also noticed
that the register bound may cause register spilling and increase the percentage
of stalls caused by memory instructions. The MemInst Stall of Im2Col+Upsample
is 43.0\% on 1080Ti without a bound, and the number increases to 73.6\% when
the kernel is launched with the bound. Because of the cost of spilled
registers, for this case the version without the register bound runs faster.
Fortunately, {\systemname} automatically profiles two different versions to
decide whether to set the register bound or not.


\section{Related Work}

\noindent\textbf{Kernel Fusion:}
\citeauthor{fusion-2010} proposed three different strategies to fusion
including concatenating the computation of two kernels similar to the standard
vertical fusion, and the distribution of the computation among different
threads similar to horizontal fusion~\cite{fusion-2010}. However, their
proposed technique cannot handle synchronization barriers and it is not
automated. Due to these limitations, their results show that distributing
computation among different threads is the worst fusion strategy out of the
three proposed fusion strategies. In contrast, out technique do not have these
limitations. Our results show that the horizontal fusion, after appropriately
handling barriers and automatically profiling the best thread space partition
scheme, often outperforms the vertical fusion.

There is a rich set of previous work that targets automatic vertical fusion.
\citeauthor{automatic-fusion} presents a searching technique that finds a
linearized kernel with lowest memory requirement~\cite{automatic-fusion}.
\citeauthor{scalable-kernel-fusion} proposes to formalize kernel fusion as an
optimization problem~\cite{scalable-kernel-fusion}. \citeauthor{ikra} proposes
a new language, called Ikra, for efficient GPU programming that allows a
programmer to implement GPU programs of multiple reusable parallel
sections~\cite{ikra}. Ikra then fuses those parallel sections into a small
number of GPU kernels. \citeauthor{cublas-fusion} present a source-to-source
compiler that is able to automatically fuse kernels that can be expressed in
the form of map and reduce calls~\cite{cublas-fusion}. All these prior works
only consider vertical fusion, rather than the horizontal fusion proposed in
our work.

\noindent\textbf{Kernel Fusion in ML:}
Kernel fusion is a critical optimization technique for machine learning
applications. \citeauthor{kernel-fusion} demonstrates several optimization
techniques including kernel fusion that can be applied to improve the
performance of RNN, GRU, and LSTM models, although they only consider
vertical fusion and the optimization can not be
automated~\cite{kernel-fusion}. Similarly, \citeauthor{persistent-rnns}
presents a framework that fused all the kernels across different time steps
into a single kernel~\cite{persistent-rnns}. However their main purpose is to
keep weight parameters stashed in registers/cache, and enable the training of
deeper RNN networks on much larger datasets. \citeauthor{astra} proposes a
compilation and execution framework that uses measurement-driven approach to
guide the compiler to select candidate kernels that can be fused~\cite{astra}.

\noindent\textbf{ML Compilers:} Machine learning compiler stack
such as TVM~\cite{tvm} and XLA~\cite{tensorflow} provide
an end-to-end solution to speed up both training and
inference phases of machine learning models. XLA has been
targeting specifically for linear optimization. TVM
demonstrates large benefits on mobile platform and IoT
devices, but incremental gain on GPU. \citeauthor{dlvm}
presents a software stack for machine learning, which
uses a linear algebraic IR and generates GPU kernels
automatically, although no evaluation results were
given regarding its performance~\cite{dlvm}.
nGraph~\cite{nGraph} library is an open-source C++ graph compiler
for Deep Learning ecosystems.
With nGraph Library, data scientists can use their preferred
deep learning framework on any number of hardware
architectures.
Glow~\cite{glow} accepts a computation graph from deep learning frameworks,
such as PyTorch, and generates highly optimized code for machine
learning accelerators.
TASO~\cite{TASO} is a deep neural network computation graph optimizer
that fuses different matrix operators using graph substitution.
Many machine learning compilers employ vertical
fusion as their standard fusion technique to save memory operations,
but none of previous ML compilers implements automatic horizontal fusion.

\noindent\textbf{Warp Specialization:}
Singe~\cite{Singe} and CudaDMA~\cite{CudaDMA} use warp specialization
techniques to speed up domain-specific applications (e.g., chemistry for Singe
and direct memory access library for CudaDMA). Similar to horizontal fusion,
the idea of warp specialization is to allow warps in a block to perform
different tasks in parallel. Comparing to {\systemname}, Singe and CudaDMA have
more significant speed up but can only apply to specific domains.

\noindent\textbf{Multi-Application Concurrency:}
Previous work also proposes techniques to enable better multi-application
concurrency~\cite{elastic-kernels, mosaic, smk, kernel-merge}.
These systems modify the GPU runtime
and may introduce overhead.
KernelMerge~\cite{kernel-merge}
modifies the OpenCL runtime to launch and execute two kernels
concurrently, which is similar to CUDA stream parallelization.
Similarly, \citeauthor{mask}
suggests to redesign the GPU memory virtualization to
mitigate address transaction overhead while supporting
multi-application concurrency~\cite{mask}.



\label{sec:related_work}

\section{Conclusion}
\label{sec:conclusion}

Automatic horizontal fusion is an effective optimization technique that
complements the standard vertical kernel fusion and it can speedup GPU programs
in domains like deep learning and cryptocurrency mining. Our experimental
results show that the horizontal fusion can enable warp schedulers in NVIDIA
GPUs to interleave instructions from different kernels to hide instruction
latencies. It is especially beneficial to apply this technique to fuse kernels
with instructions that require different kinds of hardware resources.

\bibliographystyle{IEEEtranN}
\bibliography{references}

\begin{thebibliography}{36}
\providecommand{\natexlab}[1]{#1}
\providecommand{\url}[1]{#1}
\csname url@samestyle\endcsname
\providecommand{\newblock}{\relax}
\providecommand{\bibinfo}[2]{#2}
\providecommand{\BIBentrySTDinterwordspacing}{\spaceskip=0pt\relax}
\providecommand{\BIBentryALTinterwordstretchfactor}{4}
\providecommand{\BIBentryALTinterwordspacing}{\spaceskip=\fontdimen2\font plus
\BIBentryALTinterwordstretchfactor\fontdimen3\font minus
  \fontdimen4\font\relax}
\providecommand{\BIBforeignlanguage}[2]{{%
\expandafter\ifx\csname l@#1\endcsname\relax
\typeout{** WARNING: IEEEtranN.bst: No hyphenation pattern has been}%
\typeout{** loaded for the language `#1'. Using the pattern for}%
\typeout{** the default language instead.}%
\else
\language=\csname l@#1\endcsname
\fi
#2}}
\providecommand{\BIBdecl}{\relax}
\BIBdecl

\bibitem[Nickolls et~al.(2008)Nickolls, Buck, Garland, and Skadron]{CUDA}
\BIBentryALTinterwordspacing
J.~Nickolls, I.~Buck, M.~Garland, and K.~Skadron, ``Scalable parallel
  programming with cuda,'' \emph{Queue}, vol.~6, no.~2, pp. 40--53, Mar. 2008.
  [Online]. Available: \url{http://doi.acm.org/10.1145/1365490.1365500}
\BIBentrySTDinterwordspacing

\bibitem[mlp()]{mlperf}
``Mlperf training v0.6 results,''
  \url{https://mlperf.org/training-results-0-6/}.

\bibitem[Chen et~al.(2018)Chen, Moreau, Jiang, Zheng, Yan, Shen, Cowan, Wang,
  Hu, Ceze, Guestrin, and Krishnamurthy]{tvm}
\BIBentryALTinterwordspacing
T.~Chen, T.~Moreau, Z.~Jiang, L.~Zheng, E.~Yan, H.~Shen, M.~Cowan, L.~Wang,
  Y.~Hu, L.~Ceze, C.~Guestrin, and A.~Krishnamurthy, ``{TVM}: An automated
  end-to-end optimizing compiler for deep learning,'' in \emph{13th {USENIX}
  Symposium on Operating Systems Design and Implementation ({OSDI} 18)}.\hskip
  1em plus 0.5em minus 0.4em\relax Carlsbad, CA: {USENIX} Association, Oct.
  2018, pp. 578--594. [Online]. Available:
  \url{https://www.usenix.org/conference/osdi18/presentation/chen}
\BIBentrySTDinterwordspacing

\bibitem[Jia et~al.(2019)Jia, Padon, Thomas, Warszawski, Zaharia, and
  Aiken]{TASO}
\BIBentryALTinterwordspacing
Z.~Jia, O.~Padon, J.~Thomas, T.~Warszawski, M.~Zaharia, and A.~Aiken, ``Taso:
  Optimizing deep learning computation with automatic generation of graph
  substitutions,'' in \emph{Proceedings of the 27th ACM Symposium on Operating
  Systems Principles}, ser. SOSP '19.\hskip 1em plus 0.5em minus 0.4em\relax
  New York, NY, USA: ACM, 2019, pp. 47--62. [Online]. Available:
  \url{http://doi.acm.org/10.1145/3341301.3359630}
\BIBentrySTDinterwordspacing

\bibitem[Rotem et~al.(2018)Rotem, Fix, Abdulrasool, Deng, Dzhabarov, Hegeman,
  Levenstein, Maher, Satish, Olesen, Park, Rakhov, and Smelyanskiy]{glow}
\BIBentryALTinterwordspacing
N.~Rotem, J.~Fix, S.~Abdulrasool, S.~Deng, R.~Dzhabarov, J.~Hegeman,
  R.~Levenstein, B.~Maher, N.~Satish, J.~Olesen, J.~Park, A.~Rakhov, and
  M.~Smelyanskiy, ``Glow: Graph lowering compiler techniques for neural
  networks,'' \emph{CoRR}, vol. abs/1805.00907, 2018. [Online]. Available:
  \url{http://arxiv.org/abs/1805.00907}
\BIBentrySTDinterwordspacing

\bibitem[Boemer et~al.(2019)Boemer, Lao, Cammarota, and Wierzynski]{nGraph}
\BIBentryALTinterwordspacing
F.~Boemer, Y.~Lao, R.~Cammarota, and C.~Wierzynski, ``ngraph-he: A graph
  compiler for deep learning on homomorphically encrypted data,'' in
  \emph{Proceedings of the 16th ACM International Conference on Computing
  Frontiers}, ser. CF '19.\hskip 1em plus 0.5em minus 0.4em\relax New York, NY,
  USA: ACM, 2019, pp. 3--13. [Online]. Available:
  \url{http://doi.acm.org/10.1145/3310273.3323047}
\BIBentrySTDinterwordspacing

\bibitem[Abadi et~al.(2015)Abadi, Agarwal, Barham, Brevdo, Chen, Citro,
  Corrado, Davis, Dean, Devin, Ghemawat, Goodfellow, Harp, Irving, Isard, Jia,
  Jozefowicz, Kaiser, Kudlur, Levenberg, Man\'{e}, Monga, Moore, Murray, Olah,
  Schuster, Shlens, Steiner, Sutskever, Talwar, Tucker, Vanhoucke, Vasudevan,
  Vi\'{e}gas, Vinyals, Warden, Wattenberg, Wicke, Yu, and Zheng]{tensorflow}
\BIBentryALTinterwordspacing
M.~Abadi, A.~Agarwal, P.~Barham, E.~Brevdo, Z.~Chen, C.~Citro, G.~S. Corrado,
  A.~Davis, J.~Dean, M.~Devin, S.~Ghemawat, I.~Goodfellow, A.~Harp, G.~Irving,
  M.~Isard, Y.~Jia, R.~Jozefowicz, L.~Kaiser, M.~Kudlur, J.~Levenberg,
  D.~Man\'{e}, R.~Monga, S.~Moore, D.~Murray, C.~Olah, M.~Schuster, J.~Shlens,
  B.~Steiner, I.~Sutskever, K.~Talwar, P.~Tucker, V.~Vanhoucke, V.~Vasudevan,
  F.~Vi\'{e}gas, O.~Vinyals, P.~Warden, M.~Wattenberg, M.~Wicke, Y.~Yu, and
  X.~Zheng, ``{TensorFlow}: Large-scale machine learning on heterogeneous
  systems,'' 2015, software available from tensorflow.org. [Online]. Available:
  \url{https://www.tensorflow.org/}
\BIBentrySTDinterwordspacing

\bibitem[Wei et~al.(2017)Wei, Adve, and Schwartz]{dlvm}
R.~Wei, V.~S. Adve, and L.~Schwartz, ``Dlvm: A modern compiler infrastructure
  for deep learning systems,'' \emph{ArXiv}, vol. abs/1711.03016, 2017.

\bibitem[Appleyard et~al.(2016)Appleyard, Kocisk{\'{y}}, and
  Blunsom]{kernel-fusion}
\BIBentryALTinterwordspacing
J.~Appleyard, T.~Kocisk{\'{y}}, and P.~Blunsom, ``Optimizing performance of
  recurrent neural networks on gpus,'' \emph{CoRR}, vol. abs/1604.01946, 2016.
  [Online]. Available: \url{http://arxiv.org/abs/1604.01946}
\BIBentrySTDinterwordspacing

\bibitem[Diamos et~al.(2016)Diamos, Sengupta, Catanzaro, Chrzanowski, Coates,
  Elsen, Engel, Hannun, and Satheesh]{persistent-rnns}
\BIBentryALTinterwordspacing
G.~Diamos, S.~Sengupta, B.~Catanzaro, M.~Chrzanowski, A.~Coates, E.~Elsen,
  J.~Engel, A.~Hannun, and S.~Satheesh, ``Persistent rnns: Stashing recurrent
  weights on-chip,'' in \emph{Proceedings of the 33rd International Conference
  on International Conference on Machine Learning - Volume 48}, ser.
  ICML'16.\hskip 1em plus 0.5em minus 0.4em\relax JMLR.org, 2016, pp.
  2024--2033. [Online]. Available:
  \url{http://dl.acm.org/citation.cfm?id=3045390.3045604}
\BIBentrySTDinterwordspacing

\bibitem[{Wang} et~al.(2010){Wang}, {Lin}, and {Yi}]{fusion-2010}
G.~{Wang}, Y.~{Lin}, and W.~{Yi}, ``Kernel fusion: An effective method for
  better power efficiency on multithreaded gpu,'' in \emph{2010 IEEE/ACM Int'l
  Conference on Green Computing and Communications Int'l Conference on Cyber,
  Physical and Social Computing}, Dec 2010, pp. 344--350.

\bibitem[Fousek et~al.(2011)Fousek, Filipovi\v{c}, and
  Madzin]{automatic-fusion}
\BIBentryALTinterwordspacing
J.~Fousek, J.~Filipovi\v{c}, and M.~Madzin, ``Automatic fusions of cuda-gpu
  kernels for parallel map,'' \emph{SIGARCH Comput. Archit. News}, vol.~39,
  no.~4, pp. 98--99, Dec. 2011. [Online]. Available:
  \url{http://doi.acm.org/10.1145/2082156.2082183}
\BIBentrySTDinterwordspacing

\bibitem[{Wahib} and {Maruyama}(2014)]{scalable-kernel-fusion}
M.~{Wahib} and N.~{Maruyama}, ``Scalable kernel fusion for memory-bound gpu
  applications,'' in \emph{SC '14: Proceedings of the International Conference
  for High Performance Computing, Networking, Storage and Analysis}, Nov 2014,
  pp. 191--202.

\bibitem[Filipovi{\v{c}} et~al.(2015)Filipovi{\v{c}}, Madzin, Fousek, and
  Matyska]{cublas-fusion}
\BIBentryALTinterwordspacing
J.~Filipovi{\v{c}}, M.~Madzin, J.~Fousek, and L.~Matyska, ``Optimizing cuda
  code by kernel fusion: application on blas,'' \emph{The Journal of
  Supercomputing}, vol.~71, no.~10, pp. 3934--3957, Oct 2015. [Online].
  Available: \url{https://doi.org/10.1007/s11227-015-1483-z}
\BIBentrySTDinterwordspacing

\bibitem[Chen et~al.(2015)Chen, Li, Li, Lin, Wang, Wang, Xiao, Xu, Zhang, and
  Zhang]{mxnet}
\BIBentryALTinterwordspacing
T.~Chen, M.~Li, Y.~Li, M.~Lin, N.~Wang, M.~Wang, T.~Xiao, B.~Xu, C.~Zhang, and
  Z.~Zhang, ``Mxnet: {A} flexible and efficient machine learning library for
  heterogeneous distributed systems,'' \emph{CoRR}, vol. abs/1512.01274, 2015.
  [Online]. Available: \url{http://arxiv.org/abs/1512.01274}
\BIBentrySTDinterwordspacing

\bibitem[Paszke et~al.(2017)Paszke, Gross, Chintala, Chanan, Yang, DeVito, Lin,
  Desmaison, Antiga, and Lerer]{PyTorch}
A.~Paszke, S.~Gross, S.~Chintala, G.~Chanan, E.~Yang, Z.~DeVito, Z.~Lin,
  A.~Desmaison, L.~Antiga, and A.~Lerer, ``Automatic differentiation in
  {PyTorch},'' in \emph{NIPS Autodiff Workshop}, 2017.

\bibitem[eth()]{ethminer}
``Ethminer, ethereum miner with opencl, cuda and stratum support,''
  \url{https://github.com/ethereum-mining/ethminer}.

\bibitem[ccm()]{ccminer}
``ccminer, a cuda accelerated mining application,''
  \url{https://github.com/tpruvot/ccminer}.

\bibitem[Pas()]{Pascal}
``Nvidia tesla p100, the most advanced datacenter accelerator ever built,''
  \url{https://images.nvidia.com/content/pdf/tesla/whitepaper/pascal-architecture-whitepaper.pdf}.

\bibitem[Vol()]{Volta}
``Nvidia tesla v100 gpu architecture,''
  \url{https://images.nvidia.com/content/volta-architecture/pdf/volta-architecture-whitepaper.pdf}.

\bibitem[He et~al.(2016)He, Zhang, Ren, and Sun]{resnet}
K.~He, X.~Zhang, S.~Ren, and J.~Sun, ``Deep residual learning for image
  recognition,'' 06 2016, pp. 770--778.

\bibitem[shu()]{shuffles}
``Faster parallel reductions on kepler,''
  \url{https://devblogs.nvidia.com/faster-parallel-reductions-kepler}.

\bibitem[ptx()]{ptx}
``Parallel thread execution isa version 6.5,''
  \url{https://docs.nvidia.com/cuda/parallel-thread-execution/index.html}.

\bibitem[Cla()]{Clang-cuda}
``Clang: a c language family frontend for llvm,''
  \url{https://clang.llvm.org/}.

\bibitem[Brock et~al.(2018)Brock, Donahue, and Simonyan]{biggan}
A.~Brock, J.~Donahue, and K.~Simonyan, ``Large scale gan training for high
  fidelity natural image synthesis,'' \emph{ArXiv}, vol. abs/1809.11096, 2018.

\bibitem[Li et~al.(2019)Li, Liu, Mello, Wang, Kautz, and Yang]{uvc}
X.~Li, S.~Liu, S.~D. Mello, X.~Wang, J.~Kautz, and M.-H. Yang, ``Joint-task
  self-supervised learning for temporal correspondence,'' in \emph{NeurIPS},
  2019.

\bibitem[Wood(2014)]{ethereum}
G.~Wood, ``Ethereum: A secure decentralised generalised transaction ledger,''
  \emph{Ethereum project yellow paper}, vol. 151, pp. 1--32, 2014.

\bibitem[Springer et~al.(2017)Springer, Wauligmann, and Masuhara]{ikra}
\BIBentryALTinterwordspacing
M.~Springer, P.~Wauligmann, and H.~Masuhara, ``Modular array-based gpu
  computing in a dynamically-typed language,'' in \emph{Proceedings of the 4th
  ACM SIGPLAN International Workshop on Libraries, Languages, and Compilers for
  Array Programming}, ser. ARRAY 2017.\hskip 1em plus 0.5em minus 0.4em\relax
  New York, NY, USA: ACM, 2017, pp. 48--55. [Online]. Available:
  \url{http://doi.acm.org/10.1145/3091966.3091974}
\BIBentrySTDinterwordspacing

\bibitem[Sivathanu et~al.(2019)Sivathanu, Chugh, Singapuram, and Zhou]{astra}
\BIBentryALTinterwordspacing
M.~Sivathanu, T.~Chugh, S.~S. Singapuram, and L.~Zhou, ``Astra: Exploiting
  predictability to optimize deep learning,'' in \emph{Proceedings of the
  Twenty-Fourth International Conference on Architectural Support for
  Programming Languages and Operating Systems}, ser. ASPLOS '19.\hskip 1em plus
  0.5em minus 0.4em\relax New York, NY, USA: ACM, 2019, pp. 909--923. [Online].
  Available: \url{http://doi.acm.org/10.1145/3297858.3304072}
\BIBentrySTDinterwordspacing

\bibitem[Bauer et~al.(2014)Bauer, Treichler, and Aiken]{Singe}
\BIBentryALTinterwordspacing
M.~Bauer, S.~Treichler, and A.~Aiken, ``Singe: Leveraging warp specialization
  for high performance on gpus,'' in \emph{Proceedings of the 19th ACM SIGPLAN
  Symposium on Principles and Practice of Parallel Programming}, ser. PPoPP
  '14.\hskip 1em plus 0.5em minus 0.4em\relax New York, NY, USA: ACM, 2014, pp.
  119--130. [Online]. Available:
  \url{http://doi.acm.org/10.1145/2555243.2555258}
\BIBentrySTDinterwordspacing

\bibitem[{Bauer} et~al.(2011){Bauer}, {Cook}, and {Khailany}]{CudaDMA}
M.~{Bauer}, H.~{Cook}, and B.~{Khailany}, ``Cudadma: Optimizing gpu memory
  bandwidth via warp specialization,'' in \emph{SC '11: Proceedings of 2011
  International Conference for High Performance Computing, Networking, Storage
  and Analysis}, 2011, pp. 1--11.

\bibitem[Pai et~al.(2013)Pai, Thazhuthaveetil, and
  Govindarajan]{elastic-kernels}
\BIBentryALTinterwordspacing
S.~Pai, M.~J. Thazhuthaveetil, and R.~Govindarajan, ``Improving gpgpu
  concurrency with elastic kernels,'' in \emph{Proceedings of the Eighteenth
  International Conference on Architectural Support for Programming Languages
  and Operating Systems}, ser. ASPLOS '13.\hskip 1em plus 0.5em minus
  0.4em\relax New York, NY, USA: ACM, 2013, pp. 407--418. [Online]. Available:
  \url{http://doi.acm.org/10.1145/2451116.2451160}
\BIBentrySTDinterwordspacing

\bibitem[{Ausavarungnirun} et~al.(2017){Ausavarungnirun}, {Landgraf}, {Miller},
  {Ghose}, {Gandhi}, {Rossbach}, and {Mutlu}]{mosaic}
R.~{Ausavarungnirun}, J.~{Landgraf}, V.~{Miller}, S.~{Ghose}, J.~{Gandhi},
  C.~J. {Rossbach}, and O.~{Mutlu}, ``Mosaic: A gpu memory manager with
  application-transparent support for multiple page sizes,'' in \emph{2017 50th
  Annual IEEE/ACM International Symposium on Microarchitecture (MICRO)}, Oct
  2017, pp. 136--150.

\bibitem[{Wang} et~al.(2016){Wang}, {Yang}, {Melhem}, {Childers}, {Zhang}, and
  {Guo}]{smk}
Z.~{Wang}, J.~{Yang}, R.~{Melhem}, B.~{Childers}, Y.~{Zhang}, and M.~{Guo},
  ``Simultaneous multikernel gpu: Multi-tasking throughput processors via
  fine-grained sharing,'' in \emph{2016 IEEE International Symposium on High
  Performance Computer Architecture (HPCA)}, March 2016, pp. 358--369.

\bibitem[Gregg et~al.(2012)Gregg, Dorn, Hazelwood, and Skadron]{kernel-merge}
\BIBentryALTinterwordspacing
C.~Gregg, J.~Dorn, K.~Hazelwood, and K.~Skadron, ``Fine-grained resource
  sharing for concurrent gpgpu kernels,'' in \emph{Proceedings of the 4th
  USENIX Conference on Hot Topics in Parallelism}, ser. HotPar'12.\hskip 1em
  plus 0.5em minus 0.4em\relax Berkeley, CA, USA: USENIX Association, 2012, pp.
  10--10. [Online]. Available:
  \url{http://dl.acm.org/citation.cfm?id=2342788.2342798}
\BIBentrySTDinterwordspacing

\bibitem[Ausavarungnirun et~al.(2018)Ausavarungnirun, Miller, Landgraf, Ghose,
  Gandhi, Jog, Rossbach, and Mutlu]{mask}
\BIBentryALTinterwordspacing
R.~Ausavarungnirun, V.~Miller, J.~Landgraf, S.~Ghose, J.~Gandhi, A.~Jog, C.~J.
  Rossbach, and O.~Mutlu, ``Mask: Redesigning the gpu memory hierarchy to
  support multi-application concurrency,'' \emph{SIGPLAN Not.}, vol.~53, no.~2,
  pp. 503--518, Mar. 2018. [Online]. Available:
  \url{http://doi.acm.org/10.1145/3296957.3173169}
\BIBentrySTDinterwordspacing

\end{thebibliography}

\end{document}